\documentclass[pra,showpacs,twocolumn,superscriptaddress,amsmath,amssymb]{revtex4}

\pdfoutput=1

\usepackage{graphicx}
\usepackage{color}

\usepackage{dcolumn,amsmath}
 
\begin{document}

\title{Magnetic control of ultra-cold  $^6$Li and $^{174}$Yb($^3P_2$) atom mixtures with Feshbach resonances}

\author{Alexander Petrov}
\altaffiliation{Alternative address: St. Petersburg Nuclear Physics Institute, Gatchina, 188300; Division of Quantum Mechanics, St. Petersburg State 
University, 198904, Russia.}
\author{Constantinos Makrides}
\author{Svetlana Kotochigova}
\email[Corresponding author: ]{skotoch@temple.edu}
\affiliation{Department of Physics, Temple University, Philadelphia, Pennsylvania 19122, USA}

\begin{abstract} 
We theoretically evaluate the feasibility to form magnetically-tunable 
Feshbach molecules in collisions between fermionic $^6$Li atoms and bosonic 
metastable $^{174}$Yb($^3$P$_2$) atoms. In contrast to the well-studied alkali-metal 
atom collisions, collisions with meta-stable atoms are highly anisotropic.
Our first-principle coupled-channel calculation of these collisions reveals the existence
of broad Feshbach resonances due to the combined effect of anisotropic-molecular and atomic-hyperfine interactions.
In order to fit our predictions to the specific positions of  experimentally-observed 
broad resonance structures \cite{Deep2015} we  optimized the shape of the short-range
potentials by direct least-square fitting.  
This allowed us to identify the dominant resonance by its leading angular momentum quantum numbers
and describe the role of collisional anisotropy in the creation and broadening of this and other resonances. 
\end{abstract}

\maketitle

The formation of ultracold molecules with a single unpaired
electron allowing for coupling of the electron spin with the
rotational angular momentum of the molecule has received increasing
interest \cite{Zoller2006,Carr2009,Krems2010}. For example, the authors
in Ref.~\cite{Zoller2006} proposed that such spin-rotational splitting of
$^2\Sigma^+$ molecular rotational states can make the long-range electric
dipole-dipole interaction between molecules spin-dependent. Then, these
molecules confined in two-dimensional optical lattices create a class
of ultracold molecules that can realize spin-lattice models with unique
topological properties.

Ultracold $^2\Sigma^+$ molecules can be formed from an alkali-metal
and an alkaline-earth-like atom in the ground or even long-lived
metastable states.  Experimental efforts toward production
of these ground-state molecules are in their initial stages
\cite{Okava2009,Nemitz2009,Ivanov2011,Hara2011,Khramov2014,Deep2015}.  The success of these
experiments significantly depends on the realization of a two-step
process, when a mixed quantum gas of alkali-metal and  alkaline-earth-like
atoms is first optically associated into  weakly-bound molecules, which
are then  transferred into the rovibrational ground state.

Photoassociation for homonuclear alkaline-earth-like
molecules  via  so-called optical Feshbach tuning was
pioneered by \cite{Ciurylo2005,JonesReview2006,Enomoto2008}. This
photoassociative tuning becomes possible due to the existence of
long-lived excited molecular states near the narrow intercombination
lines of the alkaline-earth-like atoms. Recent successful experiments
\cite{Yamazaki2010,Blatt2011,Yan2013, Yamazaki2013} showed that a
two-photon optical Feshbach resonance can be used to couple two colliding
atoms to a vibrational level of the molecular ground state. The suppressed
excited-state spontaneous decay makes efficient coherent
molecular formation possible.

An alternative method to form ultracold molecules from ultracold atoms
is by magneto-association.  Magnetic Feshbach resonances have had
an enormous impact on the field of laser-cooled ultra-cold atoms and
molecules \cite{Tie93,Ketterle98,Kohler2006,Chin2010,Kotochigova2014}.
These resonances have allowed a tunable, variable interaction strength
by simply varying the strength of an external magnetic field (typically
on the order of 100 G or 10 mT.) They are now widely used to investigate
Efimov physics \cite {Kraemer2006}, create ultracold molecular gasses
\cite{Kohler2006,Science08}, and to simulate unique many-body phases
\cite{Bloch2008}.  So far, these resonances have been restricted to
systems with ground-state alkali-metal and ground-state rare-earth species
\cite{Chin2010,Tiesinga2005,Petrov2012,BLev2014,Ferlaino2012B,AFrisch2013}.
Only recently have experiments searching for resonances in collisions
with electronic excited atoms sprung up. References \cite{Takahashi2013,Deep2015}
detected the first resonances between ultra-cold ground state Yb or Li and
excited Yb$^*$ in the metastable $^3$P$_2$ state.

The nature of Feshbach resonances in alkali and alkaline-earth-like atomic collisions
significantly differs from that in alkali-metal atom collisions.
In alkali metals  the hyperfine interaction between electron and
nuclear spins gives  sufficient complexity that leads to the appearance
of Feshbach resonances. In alkaline-earth-like atoms such hyperfine
interaction does not exist as the nuclear spin is decoupled from the electronic degree of freedom. 
Nevertheless, Refs.~\cite{Zuchowski2010,Brue2012}  showed that magnetically
tunable Feshbach resonances in such systems can occur and are due to
a weak $R$-dependent behavior of the hyperfine coupling constant of
the alkali-metal atom, where $R$ is the interatomic separation between
the atoms.  These resonances are predicted to be narrow, on the order
of mG, and most likely appear at  large magnetic field strengths. 
Hence, they are  difficult to observe and control.

An interesting approach to induce and modify the position and width of narrow Feshbach resonances was proposed in Ref.~\cite{Tomza2014}. 
The authors show that intense nonresonant radiation can increase the resonance widths by three orders
of magnitude up to a few Gauss. These ac-field-induced resonances can be used for the production of ultra-cold 
ground state molecules from the colliding atoms. In addition, Refs.~\cite{Marcelis2008,Li2007,Krems2006}  suggested to apply 
an external dc-electric field with or without a magnetic field to control the interaction between ultra-cold 
atoms.

Another promising way to observe broader and stronger magnetic Feshbach
resonances and subject of this study is to consider interactions between
one ground and one long-lived metastable atom.  These metastable Feshbach
resonances and its associated weakly-bound metastable molecule might be
used to efficiently transfer colliding atoms to a vibrational level of
the absolute molecular ground state.  

We report on widening the search for Feshbach resonances in
these exotic collisions by investigating heteronuclear collisions of
ground-state alkali-metal $^6$Li and  long-lived metastable rare-earth
$^{174}$Yb atoms. We  demonstrate that the existence
of magnetically-tunable Feshbach molecules for this system leads to
formation of broad and strong resonances.  Furthermore, we show that
the origin of these Feshbach resonances is the result of a combination
of exchange interactions and  scattering anisotropies, where the
interactions depend on the orientation of the interatomic axis relative
to the magnetic field direction.  The broad magnetic Feshbach molecules
have non-zero orbital angular momentum $\vec \ell$.  

Our results confirm earlier predictions by Ref.~\cite{Hutson2013} that resonances 
are strongly suppressed due to inelastic processes to energetically lower-lying
$^3$P$_1$ and $^3$P$_0$ states.  Nevertheless, a recent experiment \cite{Deep2015} on 
the $^6$Li+$^{174}$Yb($^3$P$_2$) system has shown that resonances exists. 
Here, we have performed least-square simulations on the shape of the molecular potentials 
and reproduced their resonance spectrum. 
In addition, we identified the strongest and dominant resonance in terms of its mixture of partial waves $\ell$.

\section{Interaction potentials}

\begin{figure} 
\includegraphics[scale=0.3,trim=0 28 0 53,clip]{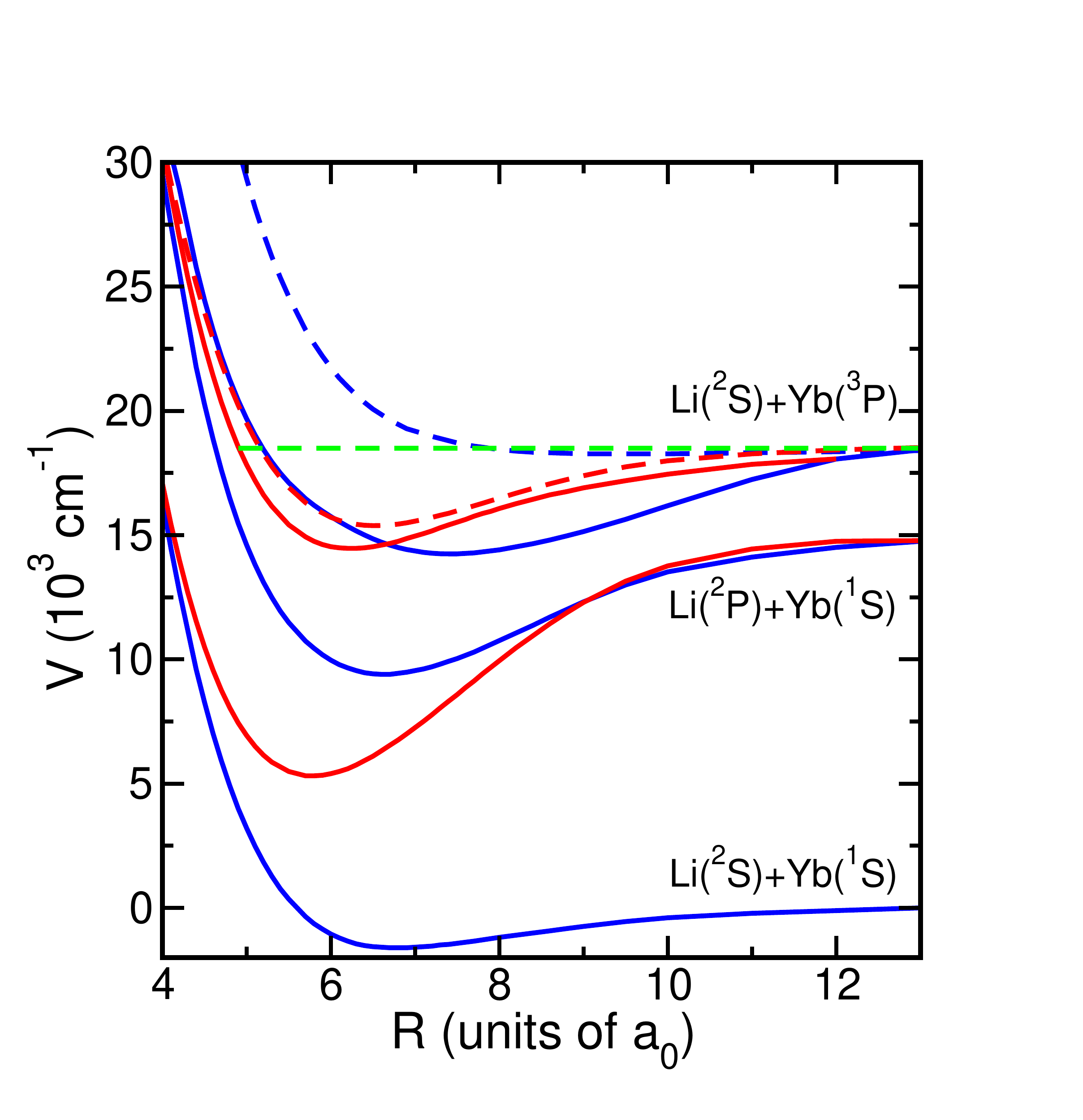} 
\caption{(Color online) The ground and lowest excited non-relativistic potentials of the LiYb
molecule as a function of internuclear separation $R$.  Here $a_0$ is
the Bohr radius of 0.0529 nm. Solid blue and red curves are $^2\Sigma^+$
and $^2\Pi$ potentials, while dashed blue and red curves are $^4\Sigma^+$
and $^4\Pi$ potentials, respectively.
Four metastable potentials dissociate to
the Li($^2$S$_{1/2}$)+Yb($^3$P) limit and are used in our coupled-channel
modeling. The green dashed horizontal line indicates the energy of
a Feshbach molecule near the $^2$S+$^3$P limit.
} 
\label{scheme} 
\end{figure}

For the determination of collisional properties of
Li($^2$S$_{1/2}$)+Yb($^3$P$_2$) system  we have ({\it i})
determined the short-range and long-range electronic interaction
potentials $V(\vec R,\tau)$ and ({\it ii}) setup a close-coupling
calculation that treats the hyperfine and Zeeman interaction,
molecular rotation, magnetic dipole-dipole interaction, and the
electronic interactions on equal footing.  Here, $\vec R$ describes
the interatomic separation and orientation and $\tau$ labels other
quantum numbers that uniquely label the electronic potentials. We have
calculated the non-relativistic  doublet $^{2}\Sigma^+$ and $^{2}\Pi$ and quartet $^{4}\Sigma^+$ and $^{4}\Pi$
potentials, shown in Fig.~\ref{scheme}, using the configuration-interaction method within
the MolPro package \cite{molpro}. 

We used the aug-cc-pCVTZ basis set  for  Li \cite{Prascher2011} and
 chose a basis set constructed from the (15s 14p 12d 11f 8g)/[8s 8p
7d 7f 5g] wave functions of M.~Dolg and X.~Cao \cite{Dolg,Dolg2013} for
Yb. The ytterbium basis  relies on a relativistic pseudopotential that
describes the  inner orbitals up to the 3d$^{10}$ shell.  Only, the 2s
valence electrons of Li and 4f$^{10}$ and 6s$^2$ valence electrons of
Yb are correlated in the {\it ab~initio} calculation.  Our potentials
agree well with the calculations of Ref.~\cite{Gopakumar2010}.  However,
the potentials are not  spectroscopically accurate and we have modified
their shape to fit to the experimental loss rate spectrum.  For $R>
27 a_0$, beyond the Le Roy radius where the electron clouds of the
atoms have negligible overlap, these electronic potentials are smoothly
connected to the long-range isotropic and anisotropic van-der-Waals
potentials determined from atomic polarizabilities \cite{Khramov2014}.
In addition, we included the magnetic dipole-dipole interaction. 

\section{Relative strength of various interactions}

\begin{figure}
\includegraphics[scale=0.3,trim=0 25 0 60,clip]{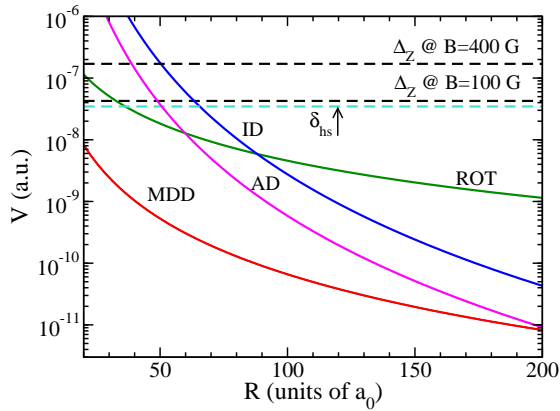}
\caption{Level splitting due to the dominant interactions between $^6$Li and Yb($^3$P) 
as a function of interatomic separation. 
Here 
         $\Delta_Z$ is the Zeeman splitting at two magnetic fields,
         $\delta_{hs}$ is hyperfine splitting of ground state $^6$Li,  
         MDD is  the magnetic dipole-dipole  interaction  $C_3/R^3$, 
         ID is the isotropic dispersion  interaction $C_6/R^6$, 
   and AD is an estimate of anisotropic dispersion  interaction  $\Delta C_6/R^6$.
   Finally, ROT is the rotational energy $\hbar^2/(2\mu_r R^2)$.
We used  $C_6=2694 E_ha_0^6 $, $\Delta C_6=242 E_ha_0^6$, and  $C_3=6.6\times10^{-5} E_ha_0^3$. 
Here $E_h=4.36 10^{-18}$ J is the Hartree.
}
\label{All}
\end{figure}

Next, we find it instructive to analyze the relative strength of
the long-range atom-atom interactions with the Zeeman, hyperfine,
and relative rotational interactions.  Figure~\ref{All} shows the two
major anisotropic interactions, the magnetic dipole-dipole $C_3/R^3$
(MDD) and the anisotropic dispersion $\Delta C_6/R^6$ (AD) interaction,
which can lead to the reorientation of the Li and Yb$^*$ angular momenta
during a collision, as a function of $R$. These potentials as well
as the isotropic dispersion $C_6/R^6$ (ID) interaction were drawn,
assuming weighted-average values of $C_3$ and $C_6$, and  $\Delta
C_6$, based on our previous calculations \cite{Khramov2014}. In fact,
$C_6=(C_{6\Sigma}+2C_{6\Pi})/3$, where $C_{6\Sigma}$ and $C_{6\Sigma}$
are the dispersion coefficient of the $\Sigma$ and $\Pi$ potentials,
respectively.  The  $\Delta C_6$ coefficient is found from tensor
analyses of the parallel and perpendicular components of the dynamic
polarizability, as described in Section 2 of Ref.~\cite{Kotochigova2010}.
In addition, the molecule can rotate, here estimated by $\hbar^2/(2\mu_r
R^2)$ with reduced mass $\mu_r$.  Finally,  hyperfine splitting (hs) of
the $^6$Li atom, and the Zeeman interaction at magnetic field  strengths
$B=100$ G and 400 G are shown.  At much shorter separation and larger
energies, not shown in the figure, the exchange interaction due to  the
exponentially-increasing electron cloud overlap lifts the degeneracy
of the double and quartet potentials and mixes the spin-orbit states of
the Yb($^3$P$_j$) levels.

The figure is most easily interpreted in the coordinate system and
angular-momentum basis set with projection quantum numbers defined along
the external magnetic field direction.  In this coordinate system the
rotational, spin-orbit, and hyperfine, and Zeeman interactions as well
as the isotropic dispersion potential shift molecular levels, whereas
the exchange interaction, the magnetic dipole-dipole interaction, and
the anisotropic component of the dispersion potential lead to coupling
between  rotational, hyperfine, and Zeeman components.  For different
interatomic separations different forces dominate.  For example, when
the curves for the magnetic dipole or anisotropic dispersion interaction
cross the hyperfine, Zeeman and/or rotational energies spin flips combined
with changes in the rotational state can occur.  In fact, the anisotropic
dispersion curve crosses the  $B$ =100 G and 400 G Zeeman curves near $R
= 50 a_0$ and $40 a_0$, respectively, whereas the magnetic dipole-dipole
curve crosses $\delta_{\rm hs}$ and $\Delta_Z$ at much shorter $R$, where
chemical bonding or exchange interaction will also play an important role.

\section{Close-coupling calculation}

We have setup a close-coupling  model for the scattering of fermionic
$^6$Li and the metastable bosonic $^{174}$Yb isotope.  The lithium atom
is uniquely specified by electron spin $s_{\rm Li}=1/2$ and nuclear
spin $i_{\rm Li}=1$.  The metastable Yb$^*$ is specified by electron
spin $s_{\rm Yb}=1$ and electron orbital angular momentum $l_{\rm Yb}=1$.
It is also convenient to define total atomic angular momenta 
${\vec f}_{\rm Li}={\vec s}_{\rm Li}+{\vec\imath}_{\rm Li}$ 
and ${\vec \jmath}_{\rm Yb}={\vec s}_{\rm Yb}+{\vec l}_{\rm Yb}$
for Li and Yb$^*$, respectively. Their projections along the $B$-field 
direction are $m_{\rm Li}$ and $m_{\rm Yb}$.
The Hamiltonian for the relative motion of the two such atoms in a magnetic field $B$ along the 
$\hat z$ direction is
\begin{equation}
H = -\frac{\hbar^2}{2\mu_r}\frac{d^2}{dR^2} + 
      \frac{\vec \ell^2}{2\mu_rR^2} + H_{\rm SO}+ H_{\rm hf}+ H_Z +U(\vec R)  \,,
\label{ham}
\end{equation}
where the rotational Hamiltonian is $\hbar^2{\vec \ell}^2/(2\mu R^2)$,
the spin-orbit interaction of Yb$^*$ is  $H_{\rm SO}=a_{\rm SO} {\vec l}_{\rm Yb}\cdot {\vec s}_{\rm Yb}$,
the hyperfine interaction of $^6$Li is  $H_{\rm hf}=a_{\rm hf} {\vec s}_{\rm Li}\cdot {\vec \imath}_{\rm Li}$,
and the Zeeman interaction 
$H_Z= (g_{e,\rm Li}s_{{\rm Li},z}+g_{N,\rm Li}i_{{\rm Li},z}+ l_{{\rm Yb},z}  + g_{e,\rm Yb} s_{{\rm Yb},z})\mu_B B$, 
with Bohr magneton $\mu_B$ and projection operators $s_{{\rm Li},z}$
etc.  The strengths $a_{\rm SO} $ and $a_{\rm hf}$ are taken from
\cite{Meggers1978,Arimondo1977}.  The  $g_{e,\rm Li}$, $g_{e,\rm Yb}$,
and $g_{N,{\rm Li}}$ are the electronic and nuclear g-factors  of Li and
Yb*  from \cite{NIST,Arimondo1977}.  Finally, $U(\vec R)$ describes the
magnetic dipole-dipole interaction and the non-relativistic electronic
potentials, which are reexpressed in terms of (tensor) coupling operators
between electron spins ${\vec s}_{\rm Li}$ and  ${\vec s}_{\rm Yb}$, and
the angular momenta ${\vec l}_{\rm Yb}$ and $\vec\ell$.  For $R\to\infty$
the interaction $U(\vec R)\to0$.

There are four contributions to $U(\vec R)$: a spin-independent isotropic potential $V_{\rm
iso}(R)$, which is proportional to an attractive $1/R^6$ potential for large separations,
a short-range isotropic exchange interaction  
$V^{\Sigma,\Pi}_{\rm exc}(R)[{\vec s}_{Li} \otimes {\vec s}_{Yb} ]_{00}=-V^{\Sigma,\Pi}_{\rm
exc}(R){\vec s}_{Li} \cdot {\vec s}_{Yb}/\sqrt{3}$, which splits doublet
from quartet $\Sigma$ and $\Pi$ potentials and  falls of exponentially
for large $R$, and an anisotropic quadrupole-like  interaction $V_{\rm ani}(R)
[\hat{C}_{2}(\hat R) \otimes [{\vec l}_{Yb} \otimes {\vec l}_{Yb}]_{2}]_{00}$,
which lifts the degeneracy of the $\Sigma$ and $\Pi$ potentials
and $V_{\rm ani}(R)\propto 1/R^6$ for large $R$. Here,  ${\hat
C}_{kq}(\hat R)$ is a spherical harmonic.
The four interaction strengths $V_{\rm iso}(R)$, $V^{\Sigma,\Pi}_{\rm exc}(R)$, and
$V_{\rm ani}(R)$ are constructed such that in the body-fixed frame
with projections along the internuclear axis the sum of the interactions closely reproduces our four
non-relativistic potentials.

The Hamiltonian is constructed in the atomic
basis or channels $Y_{\ell m_\ell}(\theta,\phi)|\alpha_{\rm Li} \rangle| \alpha_{\rm Yb} \rangle$,
where $Y_{\ell m_\ell}(\theta,\phi)$ is a spherical harmonic and  angles $\theta$
and $\phi$ give the orientation of the internuclear axis relative to
the magnetic field direction.  Hence, the rotational Hamiltonian 
is diagonal in this basis. The kets $|\alpha_{\rm Li}\rangle$ are eigen states of 
the atomic Zeeman plus hyperfine Hamiltonian of $^6$Li. For fields $B>100$ G, of interest 
in this paper and where the Paschen-Back limit holds,
we denote states by $\alpha_{\rm Li}=m_{s},m_i$, where $m_{s}$ and $m_i$ are the projections
of ${\vec s}_{\rm Li}$ and ${\vec\imath}_{\rm Li}$, respectively (and $m_{\rm Li}=m_{s}+m_i$).
The  kets $|\alpha_{\rm Yb}\rangle$ are
eigen states of the atomic spin-orbit and Zeeman Hamiltonian of $^{174}$Yb in the $^3$P term.
As the spin-orbit interaction is orders of magnitude larger than the Zeeman interaction
we denote the states by $\alpha_{\rm Yb}= j_{\rm Yb}m_{\rm Yb}$.
Coupling between the basis states is due to $U(\vec R,\tau)$.  The Hamiltonian  conserves
$M_{\rm tot} = m_{\rm Li} + m_{\rm Yb} + m_\ell$ and only even (odd) $\ell$ are coupled.

\begin{figure}
\includegraphics[scale=0.28,trim=0 25 0 0,clip]{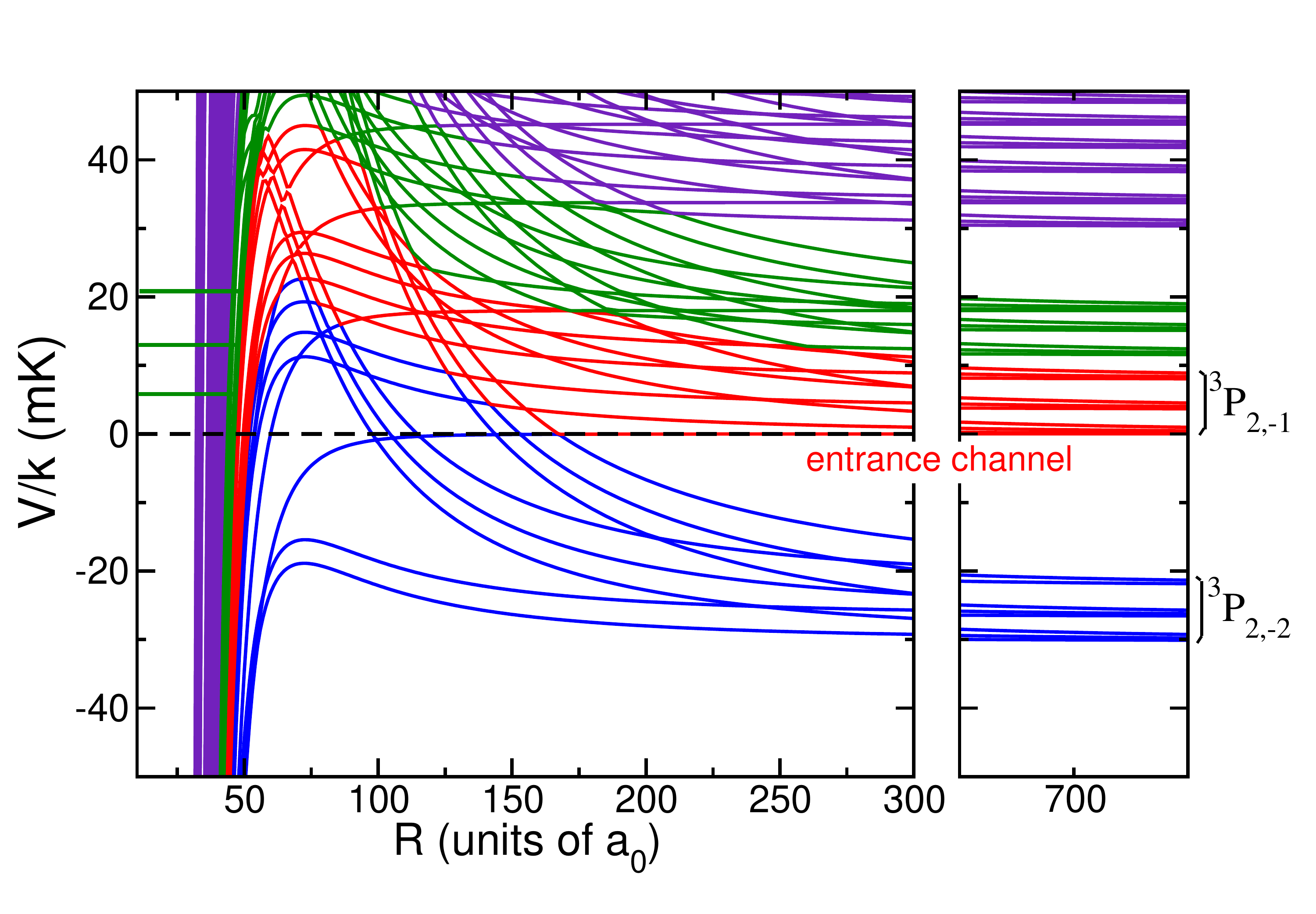}
\caption{Adiabatic potentials of  colliding $^{6}$Li($^2$S$_{1/2}$) and 
$^{174}$Yb($^3$P) near the $^3$P$_2$ dissociation limit at a magnetic field of $B=300$ G 
and $M_{\rm tot}=-1/2$ as a function of interatomic separation $R$. 
Channels with even partial wave $\ell$ up to 6 are included. 
Curves approaching the same dissociation limit correspond  to different partial waves.  
The collision starts in the $s$-wave $m_{\rm Li}=1/2$
and $j_{\rm Yb}=2$,  $m_{\rm Yb}=-1$ channel indicated by the dashed horizontal line. 
A resonance occur when the energy a bound state of a closed channel, which 
are schematically shown as dark-green horizontal solid lines close to the
entrance-channel energy. Here $k = 1.38065\cdot 10^{-23}$ J/K is the Boltzmann constant.
}
\label{Adiabats}
\end{figure}

In the experiment of Ref.~\cite{Deep2015} the $^6$Li atoms are prepared in the
energetically-lowest hyperfine state with $m_{\rm Li}=1/2$ while
the Yb atoms are prepared in the $^3$P$_2$ ($|j_{\rm Yb}m_{\rm Yb}\rangle=|2,-1\rangle$) sublevel.
For ultra-cold gasses with temperatures near 1 $\mu$K we can focus on so-called
$s$-wave or $\ell=0$ scattering. Theoretically, this corresponds to calculations including channels with $M_{\rm tot} =-1/2$
and even partial waves $\ell$. 
We will also present results for ultra-cold scattering with Yb$^*$ prepared in the state 
$|j_{\rm Yb}m_{\rm Yb}\rangle=|2,-2\rangle$. This corresponds to calculations for $M_{\rm tot} =-3/2$.

\begin{figure}
\includegraphics[scale=0.35,trim=25 -15 10 30,clip]{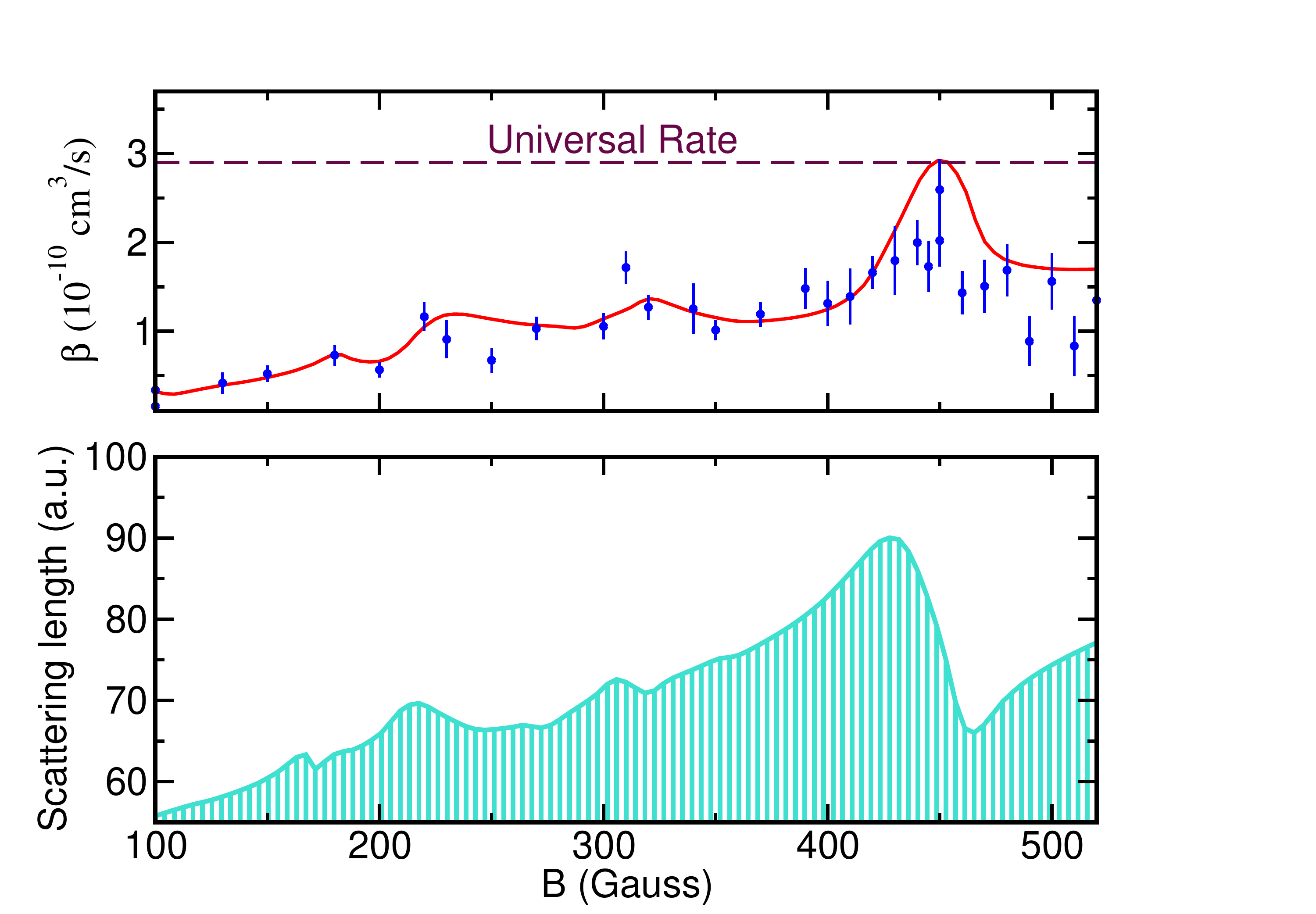} 
\caption{(Color online) Upper panel: 
Loss rate coefficient in the 
$^{6}$Li($m_s=-1/2,m_i=1$)+$^{174}$Yb($^3{\rm P}_2,j=2, m=-1$) collision as a function of magnetic field $B$.  
The experimental data (blue dots with one standard-deviation uncertainties) is obtained at a temperature 
of $T=1.6$ $\mu$K. The theoretical simulation (solid red curve)
with optimized short-range potentials is obtained for the same temperature.
Lower panel: Scattering length as a function of $B$ at collision energy $E/k=1.6$ $\mu$K. 
The short-range potentials are as the same as for the rate in upper panel.
Two dashed line indicates the van der Waals length  
with average value of $C_6$=2694 a.u.. 
}
\label{rate}
\end{figure}

Figure~\ref{Adiabats} shows a sample of the medium- to long-range adiabatic potentials, obtained by diagonalizing $H$ excluding 
the radial kinetic energy operator, as a function of interatomic separation
$R$ near the Yb$^*$($^3$P$_2$) dissociation limit at a magnetic
field of $B=300$ G. Only potentials dissociating to the
$f_{\rm Li}=1/2$ or $3/2$ and $j_{\rm Yb}=2$,  $m_{\rm Yb}=0, -1,-2$
limits are visible in the figure.  Inelastic collisional losses to
energetically lower-lying  Yb$^*$($^3$P, $j_{\rm Yb}=0,1$) states do nevertheless exist.
Feshbach resonances occur in channels with a dissociation energy above
that of the entrance channel.

Figure \ref{rate} shows the experimental  inelastic rate coefficient \cite{Deep2015} at 
a temperature of $1.6\, \mu$K in the collision between  $^{6}$Li
and $^{174}$Yb$^*$ atoms and our best-fit
theoretical inelastic rate coefficient  at a collision energy of $E/k=1.6\, \mu$K.
We include channels with $M_{\rm tot}=-1/2$ and  even $\ell$ up to $8$
leading to a close-coupling calculation with 99 channels. A thermal
average of the theoretical inelastic rate coefficient is not required as we are in the Wigner threshold limit
where this rate is independent of collision energy.
We observe at least one clear resonance at $B=450$
G and possibly three weaker resonances. From calculations that include fewer partial waves we  find that the
resonances can not be labeled by a single partial wave. Their locations
shift significantly and only converge to within a few Gauss when $\ell=8$ channels are included.
Including channels with $\ell\le10$ shifts resonance by no more than 0.1 G. This is well within the width 
of the resonances for both Figs. 4 and 5. Consequently, we conclude that it is sufficient to include channels up to $\ell\le8$.

Our bound-state calculation allows us to identify the dominant closed channel that  couples
to the s-wave entrance channel at $B=450$ G. We identify this channel with the $\ell$=4
partial wave and  projection $m_{\ell}$=1. This coupling is  due to  anisotropic  interactions, such 
as the anisotropic dispersion forces at large separations. Moreover, this channel dissociates to the atomic limit
$^6$Li($m_s$=1/2, $m_i$=-1) + $^{174}$Yb($^3$P, j=2, m=-1), which lies above the entrance channel (see Fig. 3). 

Figure~\ref{rate} also provides evidence that the inelastic rate of the strong resonance is close 
to the universal rate, as shown by the dashed line in Fig.~\ref{rate}. This is a clear indication 
that the collisional system is close to the universal regime for losses on the dominant resonance and lies well
below for other magnetic fields \cite{Idziaszek2010, Kotochigova2010}.

The agreement between the experimental \cite{Deep2015} and our theoretical 
spectrum is only obtained by allowing the coefficients $V_{\rm iso}(R)$, $V^{\Sigma,\Pi}_{\rm exc}(R)$,
and $V_{\rm ani}(R)$ to vary such that the depth of the four non-relativistic potential curves is changed.
We can not exclude the existence of other shapes of  potentials which will lead to a loss rates that is
consistent with the experimental data.

The lower panel of Fig.~\ref{rate} shows the scattering length $a$ as a function of the magnetic field strength for the collisions
between $^{6}$Li($m_s=-1/2,m_i=1$) and $^{174}$Yb($^3{\rm P}_2,j_{\rm
Yb}=2, m_{\rm Yb}=-1$) atoms at a collisional energy of 1.6 $\mu$K. 
Here the scattering length is defined through $S_{\rm elastic}= \exp(-2ik[a-ib])$,
where $S_{\rm elastic}$ is the elastic S-matrix element for the $s$-wave scattering channel,
wavenumber $k$ is given by $E=\hbar^2k^2/(2\mu_r)$, 
and positive length $b$ can be related to the inelastic loss rate coefficient.
The resonance features of $a$ correspond to those for the inelastic rate in Fig.~\ref{rate}. The
Feshbach resonance centered at 450 G has a well resolved Fano profile
with a background value close to $75 a_0$.  

We have also studied the role of the Yb spin-orbit interaction and anisotropy 
of the electronic potentials on the Feshbach resonance structure. Our analyses indicate
that for most magnetic fields the resonance features are broadened by the losses energetically 
lower collision channels with non-zero orbital angular momentum $\ell$. This confirmes our
predictions about the importance of anisotropic coupling between the short-range 
potentials. Moreover, we  also observe that, except for small fields, more than one half of the
loss goes into $^3$P$_2,j=2, m=-2$ channels. In addition, a significant fraction of the population 
ends up in the Yb($^3$P$_1$) spin-orbit state (see Fig.4(b) in Ref.~\cite{Deep2015}).

\begin{figure}
\includegraphics[scale=0.28,trim=0 25 0 10,clip]{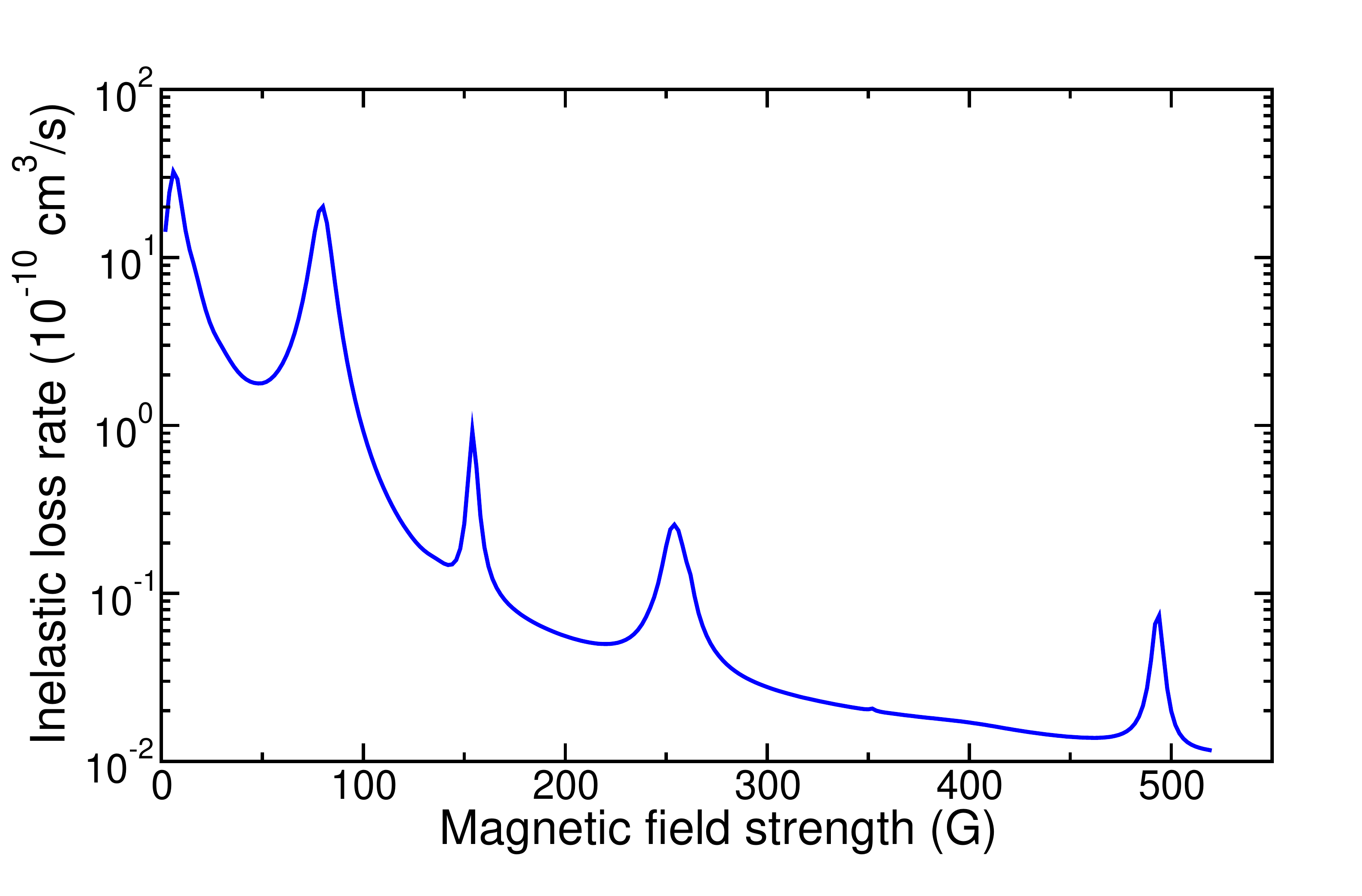}
\caption{(Color online)  Predicted theoretical loss rate coefficient in the 
$^{6}$Li($m_s=1/2,m_i=-1$)+$^{174}$Yb($^3{\rm P}_2,j_{\rm Yb}=2, m_{\rm Yb}=-2$) collision with $M_{\rm tot}=-3/2$  
and collision energy of $E/k=1.6$ $\mu$K 
as a function of magnetic field  $B$.  The  simulation is performed with optimized short-range potentials as in Fig.~\ref{rate}.
}
\label{lowestmj}
\end{figure}

\section{Conclusion}

We discussed  advances in the  understanding of scattering properties
of high-spin open-shell atomic systems. In particular, our attention
was directed towards collisions between fermionic lithium atoms in their
absolute ground state with bosonic ytterbium in its meta-stable $^3$P$_2$
state.  Both species, either with their bosonic or fermionic isotopes,
have been successfully cooled to quantum degeneracy and confined in
optical dipole traps or optical lattices, allowing the study of their
collisions at the quantum level.  We observed magnetic Feshbach resonances
in collisions with the meta-stable $^{174}$Yb atoms even though this atom
was prepared in the $m_{\rm Yb}=-1$ magnetic sub level, which is not
the energetically-lowest $^3$P$_2$ Zeeman state.  With couple-channels
calculations we have elucidated the role of anisotropies in the creation
of these resonances and by optimizing the shape of the  short-range
potentials reproduced the measured loss rate for magnetic fields up to
500 G. 

Initially, collisions between ground state Li and metastable Yb in the  $^3P_2$ state seemed to be a good system in which to form strong Feshbach resonances, 
since the Yb life time is 15 s and interactions with Li atoms is highly anisotropic. Our calculations, however, show that when Yb atoms are not prepared in the  
energetically-lowest  magnetic sublevel the  resonances have large inelastic losses (on the order of 10$^{-10}$ cm$^{3}$/s).  
This large rate coefficient might be an obstacle to use them in  stimulated Raman adiabatic passage schemes to create the ground state molecules.    
On the other hand, we also found that  inelastic loss for the resonances in collisions with  the energetically-lowest magnetic sublevel 
of Yb is an order of magnitude smaller. These resonances if experimentally confirmed can be promising candidates for
Raman transfer. In Fig.~\ref{lowestmj} we present a prediction for the Feshbach spectrum for the collision between $^{6}$Li and $^{174}$Yb ($^3{\rm
P}_2,j=2, m=-2$) based on the optimized potentials.

\section{Acknowledgments} This work is supported by an AFOSR grant No. FA9550-14-1-0321, the ARO MURI grant No. W911NF-12-1-0476,
and the NSF grant No. PHY-1005453.  

\bibliography{BibTexLibraryKotochigova}   

\end{document}